\documentclass[12pt]{iopart}
\usepackage{graphicx}  

\begin{document}
\title[Highlights from the NA60 experiment]{Highlights from the NA60 experiment}

\newcommand{\torinoI}{$^a$}
\newcommand{\torinoU}{$^b$}
\newcommand{\heid}{$^c$}
\newcommand{\bern}{$^d$}
\newcommand{\cern}{$^e$}
\newcommand{\lpc}{$^f$}
\newcommand{\llr}{$^g$}
\newcommand{\bnl}{$^h$}
\newcommand{\ipn}{$^i$}
\newcommand{\caglI}{$^j$}
\newcommand{\caglU}{$^k$}
\newcommand{\riken}{$^l$}
\newcommand{\yer}{$^m$}
\newcommand{\lis}{$^n$}

\author{Alessandro De Falco for the NA60 Collaboration\\
R~Arnaldi\torinoI,
K~Banicz\heid~\cern, 
K~Borer\bern,
J~Castor\lpc, 
B~Chaurand\llr, 
W~Chen\bnl,
C~Cicalo\caglI,  
A~Colla\torinoU~\torinoI, 
P~Cortese\torinoU,
S~Damjanovic\heid~\cern, 
A~David\cern~\lis, 
A~de~Falco\caglU~\caglI, 
A~Devaux\lpc, 
L~Ducroux\ipn, 
H~En'yo\riken,
J~Fargeix\llr, 
A~Ferretti\torinoU~\torinoI, 
M~Floris\caglU~\caglI, 
P~Force\lpc,
A~Forster\cern,
N~Guettet\cern~\lpc,
A~Guichard\ipn, 
H~Gulkanian\yer, 
J~Heuser\riken,
M~Keil\cern~\lis, 
L~Kluberg\llr, 
Z~Li\bnl, 
C~Louren\c{c}o\cern,
J~Lozano\lis, 
F~Manso\lpc, 
P~Martins\cern~\lis,  
A~Masoni\caglI,
A~Neves\lis, 
H~Ohnishi\riken, 
C~Oppedisano\torinoU~\torinoI,
P~Parracho\cern~\lis, 
P~Pillot\ipn, 
T. Poghosyan\yer,
G~Puddu\caglU~\caglI, 
E~Radermacher\cern,
P~Ramalhete\cern~\lis, 
P~Rosinsky\cern, 
E~Scomparin\torinoI,
J~Seixas\lis, 
S~Serci\caglU~\caglI, 
R~Shahoyan\lis, 
P~Sonderegger\cern~\lis,
H~J~Specht\heid, 
R~Tieulent\ipn, 
G~Usai\caglU~\caglI, 
R~Veenhof\cern,
H~W\"ohri\caglI~\lis 
}

\address{
~{\torinoI}INFN Torino,~Italy\\
~{\torinoU}Universit\`a di Torino,~Italy\\
~{\heid}Physikalisches Institut der Universit\"{a}t Heidelberg, Germany\\
~{\bern}Laboratory for High Energy Physics, University of Bern, Bern, Switzerland\\
~{\cern}CERN, 1211 Geneva 23, Switzerland\\
~{\lpc}LPC, Universit\'e Blaise Pascal and CNRS-IN2P3, Clermont-Ferrand, France\\
~{\llr}LLR, Ecole Polytechnique and CNRS-IN2P3, Palaiseau, France\\
~{\bnl}Brookhaven National Laboratory, Upton, New York, USA\\
~{\ipn}IPN-Lyon, Universit\'e Claude Bernard Lyon-I and CNRS-IN2P3, Lyon, France\\
~{\caglI}INFN Cagliari, Italy\\
~{\caglU}Universit\`a di Cagliari,~Italy\\
~{\riken}RIKEN, Wako, Saitama, Japan\\
~{\yer}YerPhI, Yerevan Physics Institute, Yerevan, Armenia\\
~{\lis}Instituto Superior T\'ecnico, Lisbon, Portugal\\
}

\ead{alessandro.de.falco@ca.infn.it}
\begin{abstract}
NA60 measured dimuon production in p-A and In-In collisions at the CERN
SPS. This paper presents a high statistics measurement of $\phi$ meson
production in In-In collisions at 158 AGeV. Both the transverse
momentum, rapidity, decay angular distributions and the absolute yield
were measured as a function of centrality. The results are compared to
previous measurements in order to shed light on the long standing $\phi$
puzzle. In addition, highlights on $\eta$ meson production and 
on the dimuon excess below the J/$\psi$ mass
are presented.
\end{abstract}

\maketitle

\section{Introduction}
Lepton pairs production is particularly suitable as a probe of the hot and dense 
medium created in high-energy heavy ion collisions, because
they can access several Quark-Gluon Plasma (QGP) signatures. 
Among them is strangeness 
enhancement~\cite{Rafelski} that can be addressed by measuring 
the $\phi$ and $\eta$ mesons production. The production of thermal 
dileptons~\cite{Shuryak}, that can be detected in the intermediate mass 
region (IMR) between the $\phi$ and the $J/\psi$ peaks, is another key 
observable. 

NA60 is an SPS third generation experiment designed to study muon pairs 
production with high precision and statistics. In the following, after a 
short description of the experimental apparatus, key results on $\phi$ and
$\eta$ production and on the excess in the intermediate mass region 
will be discussed.

\section{NA60 Detector and data selection}
NA60, a fixed target experiment at the CERN SPS, is mainly dedicated to the 
detection of muon pairs. The apparatus and data selection procedure are 
fully described in~\cite{imrNA60}. 
The key detectors 
are the muon spectrometer (MS) and the vertex tracker (VT). 
The MS detects the muon tracks in the rapidity range $3<y<4$ , and is preceded by a
12~$\lambda_I$ thick hadron absorber. The vertex tracker is a Si pixel detector 
placed between the target and the absorber in a 2.5~T dipole field, that tracks the 
muons, among the other tracks, before they suffer multiple scattering and energy loss. 
The muons detected in the MS are matched to the VT tracks both in 
momentum and angle, thus
improving the mass resolution and making it possible to measure the 
distance between the primary vertex and the track impact point.

The sample used for the results presented in this paper was collected with
a 158~A$\cdot$GeV In beam impinging on a segmented In target. 
$2.3\cdot 10^{8}$
triggers were acquired. 
In order to avoid events with reinteractions of 
nuclear fragments in the subsequent targets, only events with one vertex
in the target region were selected. 
The combinatorial background was evaluated by means of the 
event mixing technique, with an accurancy of $\sim 1\%$ 
over the full dimuon mass spectrum. 
The contribution due to the incorrect match between
the VT tracks and the muons detected in the MS was estimated either with an 
overlay Monte Carlo simulation or with a mixing technique. The former consists
in generating a dimuon on top of a real event, while the latter mixes the MS
muons of a given event with the VT tracks of another event. The two 
methods agree within $5\%$. 
The final sample consist of 440000 
signal muon pairs, with an average signal to background ratio of about 1/7. 

\section{$\phi$ production in In-In collisions}
NA60 can detect the $\phi$ meson via its decay into both muon and
kaon pairs, thus helping understanding the so-called $\phi$ puzzle. This 
takes its origin from the discrepancies observed in the $\phi$ measurements 
performed by NA50~\cite{na50phi} in the muon channel and by 
NA49~\cite{na49phi} in the kaon channel, both
in Pb-Pb collisions at 158~A$\cdot$GeV. The multiplicity measured by NA50 is 
much higher than the one obtained by NA49. 
Also the inverse slopes $T_{eff}$ of the transverse mass distributions differ: NA50 
finds a value of about 230~MeV, almost independent on centrality, while
NA49 measures values increasing with the number of participants $N_{part}$ 
from $\sim$250~MeV to $\sim$300~MeV. It has to be stressed that the acceptance
of NA50 is limited to $p_T>1.1$~GeV/$c$, while NA49 is dominated by low 
transverse momenta: in presence of radial flow, the measured $T_{eff}$ 
will depend on the fit range, being lower at high $p_T$. 

The origin of this discrepancy has been longely discussed. It was suggested  
that kaons may suffer rescattering and absorption in the medium, resulting in 
a depletion of kaon pairs, especially at low $p_T$, that leads to a reduced
yield and a hardening of the $p_T$ distribution in the hadronic channel~\cite{kaondepl}. 
Recent results by CERES~\cite{ceresphi} both in the hadronic and leptonic 
channels confirm the NA49 results, and thus suggests that there is no 
$\phi$ puzzle. However, the results in dielectrons, being affected by a 
large error bar, are not conclusive. 
 
NA60 measures the $\phi$ yield in the muon channel by counting the events in 
the mass window $0.98<M_{\mu \mu}<1.06$~GeV/$c^2$~\cite{micheleQM08}. The signal/background
ratio below the $\phi$ peak is $\sim$1/3, integrated over centrality.
To account for the continuum under the 
$\phi$ peak, two side windows between $0.88<M_{\mu \mu}<0.92$~GeV/$c^2$ 
and $1.12<M_{\mu \mu}<1.16$~GeV/$c^2$ are subtracted to the 
$\phi$ window. By performing the measurement in several $p_T$, rapidity and
decay angle intervals, raw differential distributions are extracted. 
They are then corrected for the acceptance using an overlay Monte Carlo 
tuned iteratively such that the resulting distributions reproduce the measured
data. The systematic error is evaluated by varying the analysis parameters 
and selections. The measurement is performed for 5 centrality bins, 
corresponding to $\left < N_{part} \right > =15,39,75,132,183$.

The rapidity distribution, integrated over centrality, 
is fitted with a gaussian, giving a width $\sigma=1.13 \pm 0.06 \pm 0.05$. 
No centrality dependence is observed. The result is in agreement with the 
NA49 results on several colliding systems~\cite{na49phi},\cite{na49phi2}.
The decay angle distribution is obtained for several reference systems
(Helicity, Collins-Soper, Gottfried-Jackson) resulting flat in all the cases, 
independent of centrality.  

\begin{figure}[tbh]
  \centering
  \includegraphics[width=0.427\textwidth]{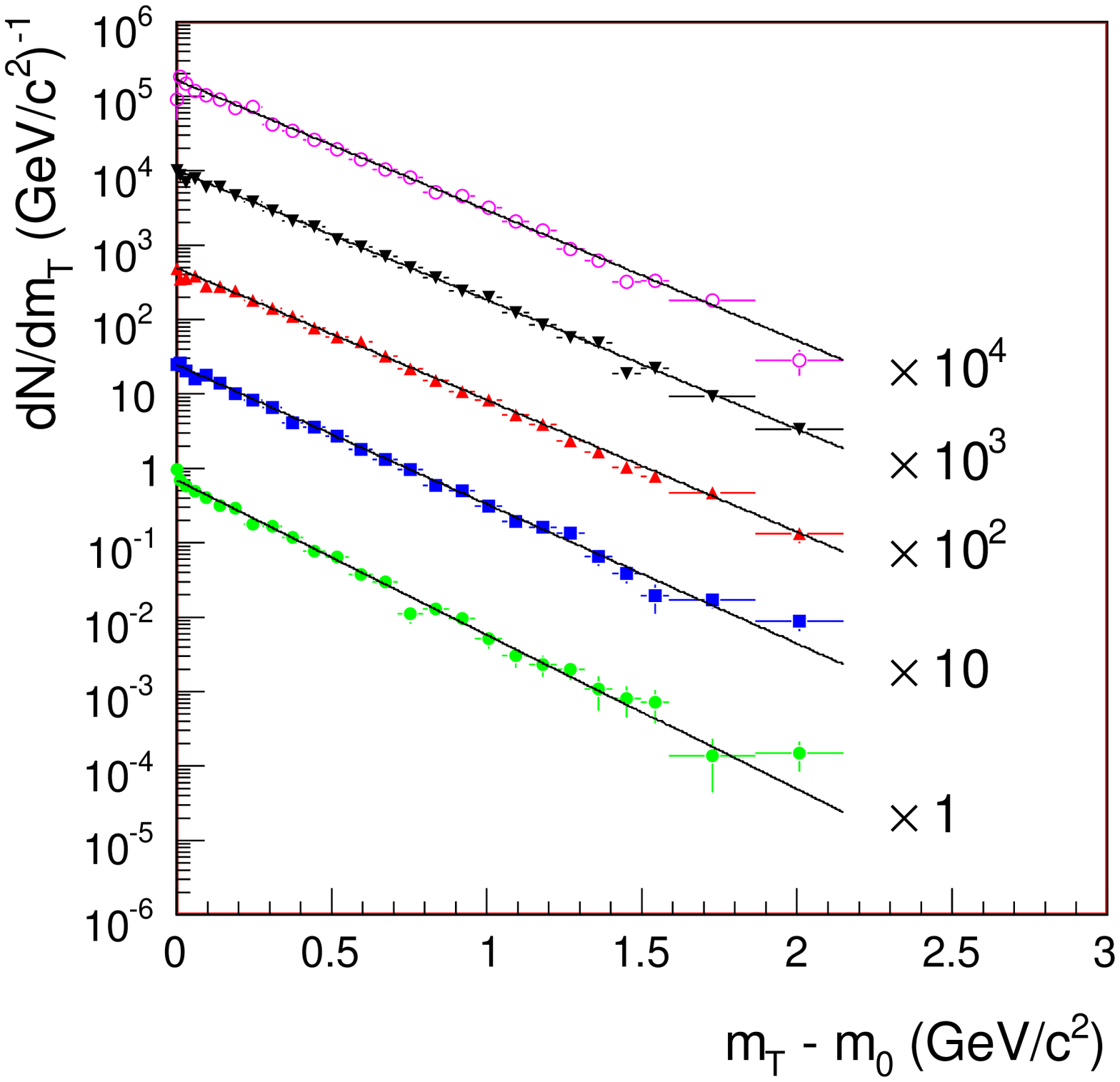}
  \includegraphics[width=0.413\textwidth]{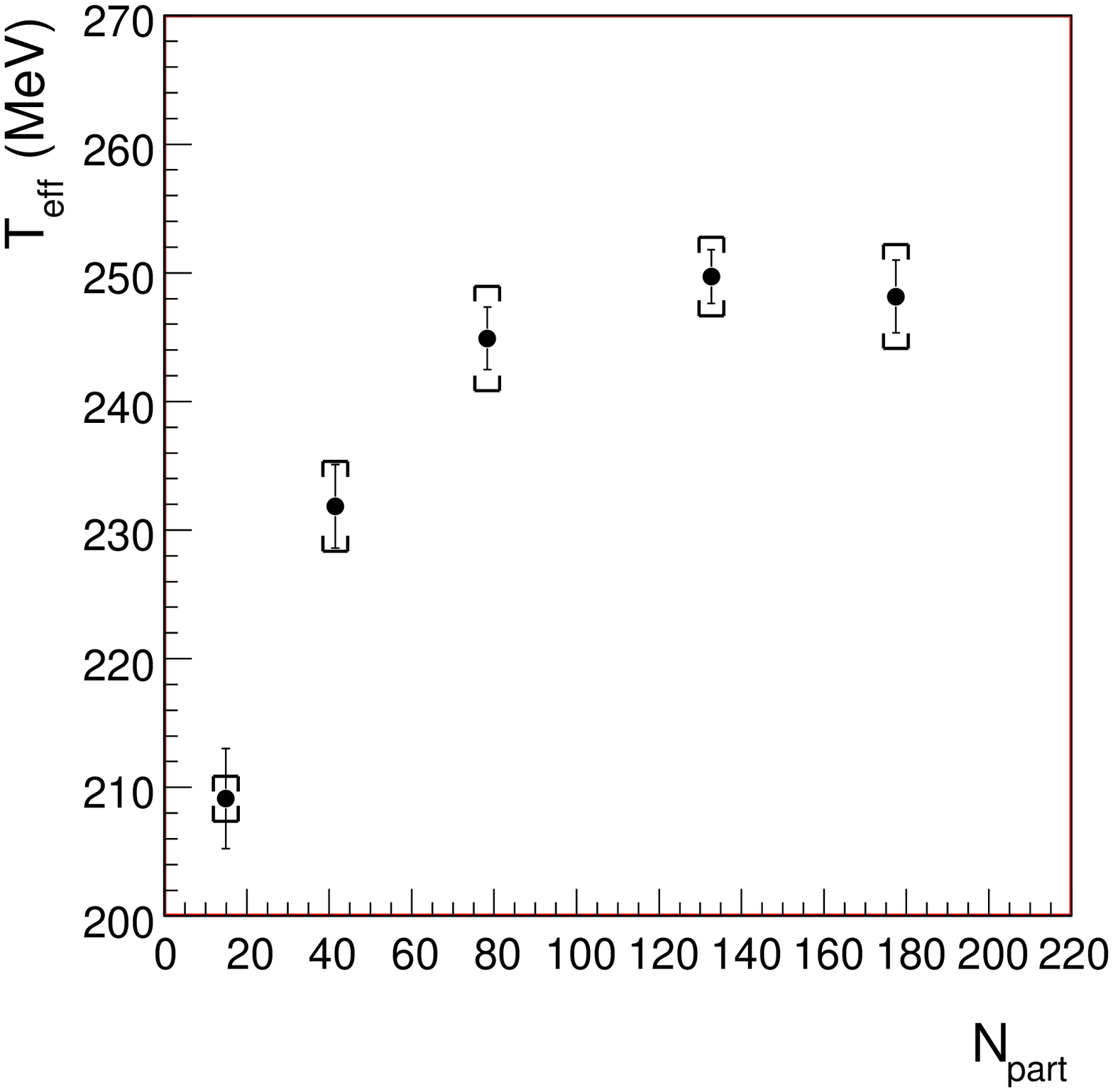}
  \caption{Left panel: $\phi$ transverse momentum distributions in
    indium-indium collisions as a function of centrality. From top to
    bottom: central to peripheral spectra.  Right Panel: $T$ slope
    parameter as a function of the number of participants. 
}
  \label{fig:mtdistr}
\end{figure}

\begin{figure}[tbh]
  \centering
  \includegraphics[width=0.41\textwidth]{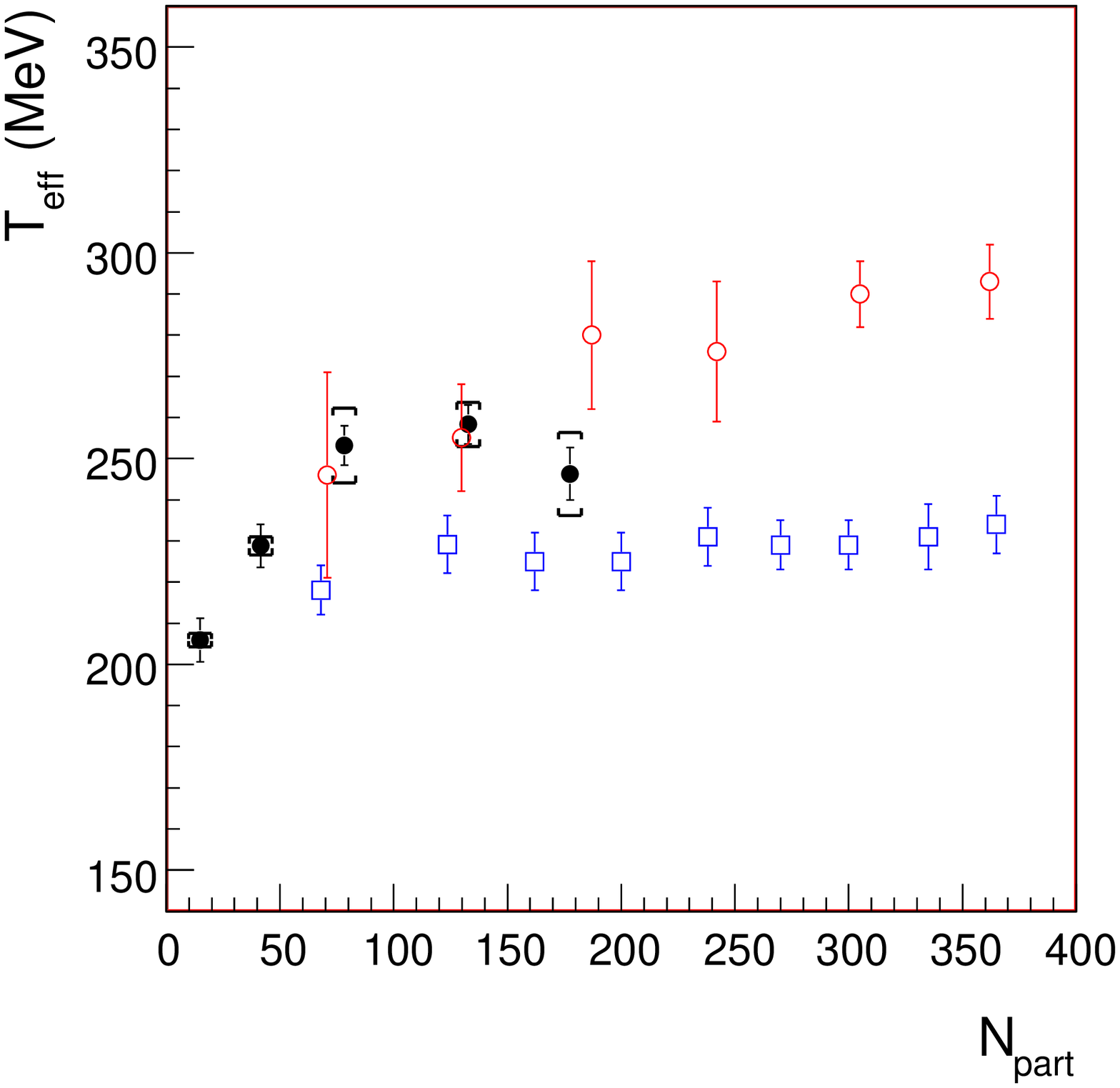}
  \includegraphics[width=0.41\textwidth]{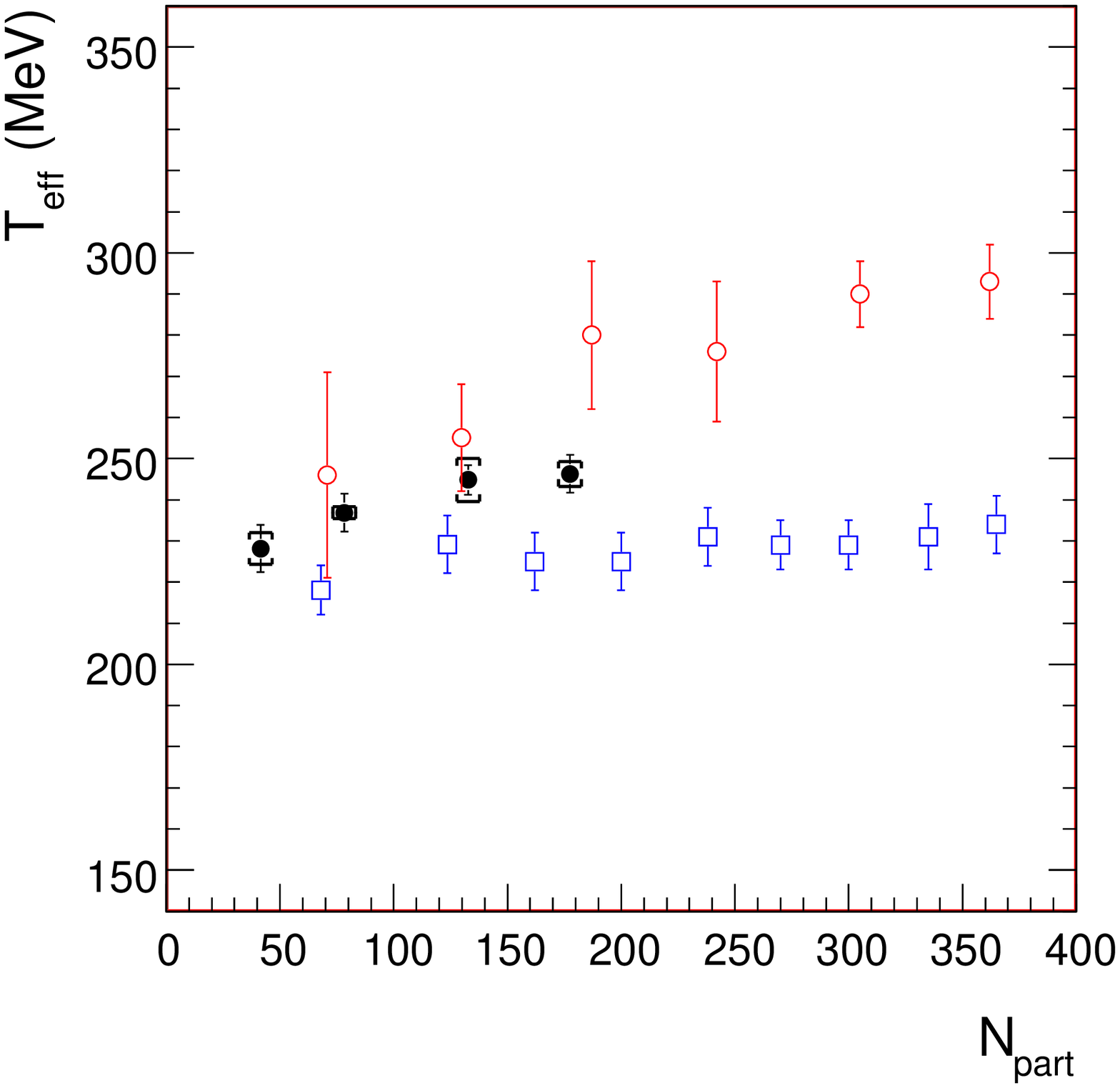}
  \caption{Centrality dependence of the $T_{eff}$ slope parameter (full
    circles) compared to NA49 (open circles) and to NA50 (open
    squares). Left panel: fit performed in the NA49 range ($0 <
    p_\mathrm{T} < 1.6~$GeV/$c$).  Right panel: fit performed in the
    NA50 range ($1.1 < p_\mathrm{T} < 3$~GeV/$c$). }
  \label{fig:TvsNA49NA50}
\end{figure}

The transverse mass distributions, shown in fig.~\ref{fig:mtdistr} (left) for 5 
centrality bins, are fitted with the function $1/m_T dN/dm_T = N_0 exp(-m_T/T_{eff})$. 
The resulting inverse slopes are plotted as a function of $N_{part}$ in 
fig.~\ref{fig:mtdistr} (right), showing an increase from peripheral to mid-central 
collisions and a saturation for most central collisions. 
Since in presence of radial flow the $T_{eff}$ value may depend on the fit range, 
in order to compare our results with NA49 and NA50, the fit is also 
performed for $p_T<1.6$~GeV/$c$ and $p_T>1.1$~GeV/$c$. The results are shown 
in fig.~\ref{fig:TvsNA49NA50} for the two $p_T$ ranges.  At low $p_{T}$ 
(fig.~\ref{fig:TvsNA49NA50}, left),  the NA60 In-In data show an increase 
with $N_{part}$, in agreement with the NA49 results in the same range. 
At high $p_T$ (fig.~\ref{fig:TvsNA49NA50}, right) a flattening of the
inverse slope dependence on centrality is observed, analogously to the NA50 
results. However, while the difference between the low-$p_T$ NA49 results and the
high-$p_T$ NA50 results amounts to $\sim$70~MeV for central collisions, the same 
difference in the NA60 In-In data is of $\sim$15~MeV. While the latter can 
presumably be due to radial flow, the magnitude of the effect in Pb-Pb can 
hardly be attributed entirely to the same cause. A further flattening due
to kaon rescattering and absorption may lead to larger $T_{eff}$ values in the 
NA49 data. 

The $\phi$  multiplicity was obtained with two different approaches, with largely
independent systematic errors. In the first, the cross section is directly measured, 
counting the number of $\phi$ produced and the number of incident ions.
The average $\phi$ multiplicity is then obtained dividing the $\phi$ cross 
section by the total In-In inelastic cross section, estimated with the general 
formula for two nuclei of mass numbers A and B:
$\sigma_{inel,AB} = \pi r^2_0 \left[A^{1/3} + B^{1/3} -\beta \left(
A^{-1/3} + B^{-1/3} \right) \right]$ ~\cite{anikina}, with
$r_0=1.3$ and $\beta=1.93$ which leads to $\sigma_{inel,In-In}=4.32$~b.
Alternatively, the $\phi$ multiplicity can be obtained using the J/$\psi$ as a 
calibration process. The ratio between the $\phi$ and the $J/\psi$ multiplicities
is given by: $\left<\phi\right>/\left<J/\psi\right> =
(N_{\phi}/A_{\phi}\times\varepsilon_{rec,\phi})/
(N_{J/\psi}/A_{J/\psi}\times\varepsilon_{rec,J/\psi})$, where 
$N_{\phi(J/\psi)}$, $A_{\phi(J/\psi)}$ and $\epsilon_{rec.\phi(J/\psi)}$ are
the number of observed $\phi~(J/\psi)$ resonances, the acceptance and
the reconstruction efficiency respectively.  Once
corrected for anomalous and nuclear absorption, $\left<J/\psi\right>$
scales with the number of binary collisions $N_{coll}$ 
and can be written as
$\left<J/\psi\right> = (\sigma_{NN}^{J/\psi}/\sigma_{NN})\cdot
N_{coll}$, where $\sigma_{NN}^{J/\psi}$ and $\sigma_{NN}$
are the $J/\psi$ and inelastic cross sections in nucleon-nucleon collisions,
and $N_{coll}$ is calculated from $N_{part}$ using the Glauber model. 
By multiplying this value by the above ratio, one obtains
$\left<\phi\right>$. 
The two methods agree within $\sim10\%$.

\begin{figure}[t]
\begin{minipage}[t]{0.49\textwidth}
  \centering
  \includegraphics[width=0.97\textwidth]{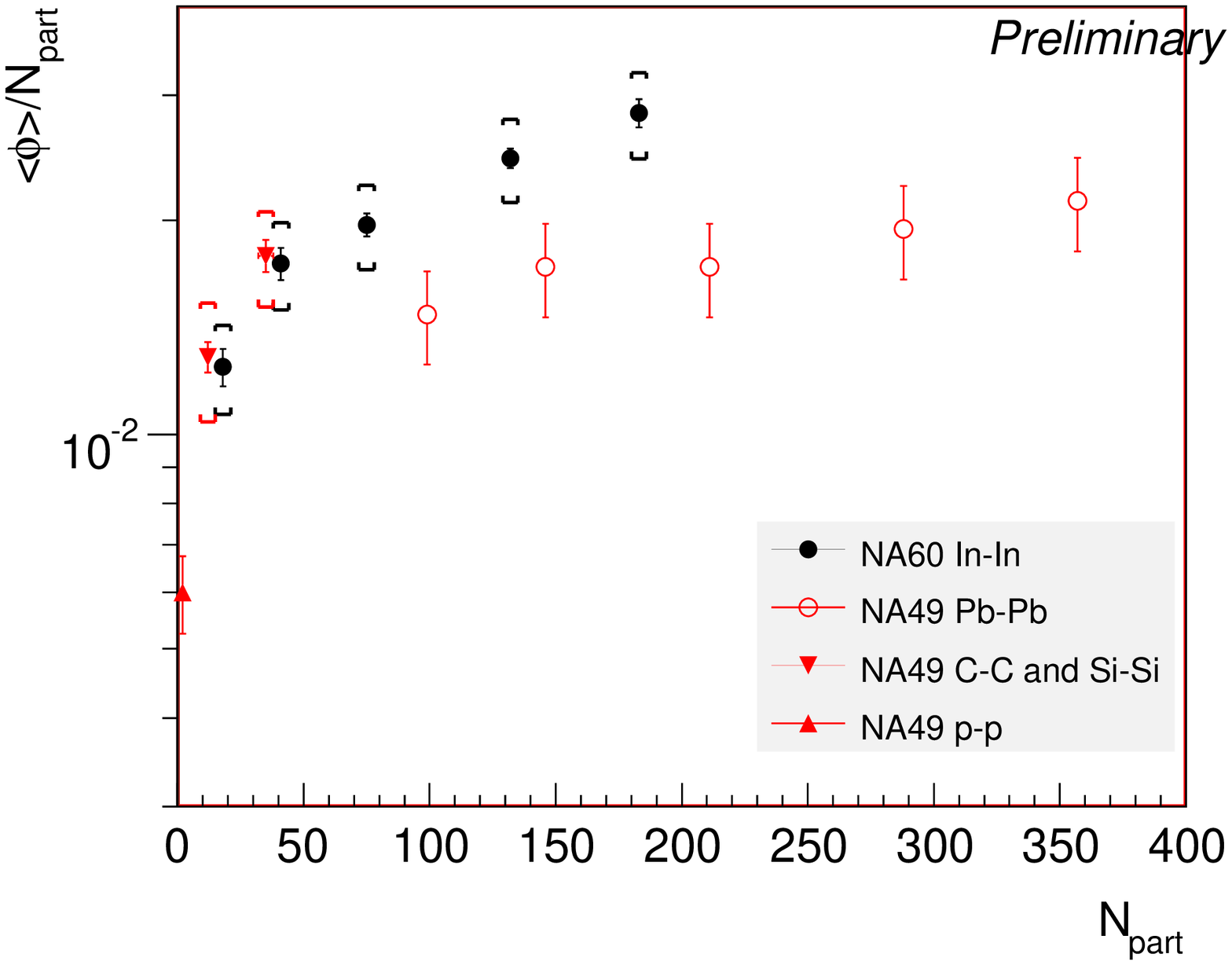}
  \caption{$\left < \phi \right >/N_{part}$ in full phase space, compared
    to the NA49 measurement in several collision systems.}
  \label{fig:phimult}
\end{minipage}
\begin{minipage}[t]{0.49\textwidth}
  \centering
  \includegraphics[width=0.95\textwidth]{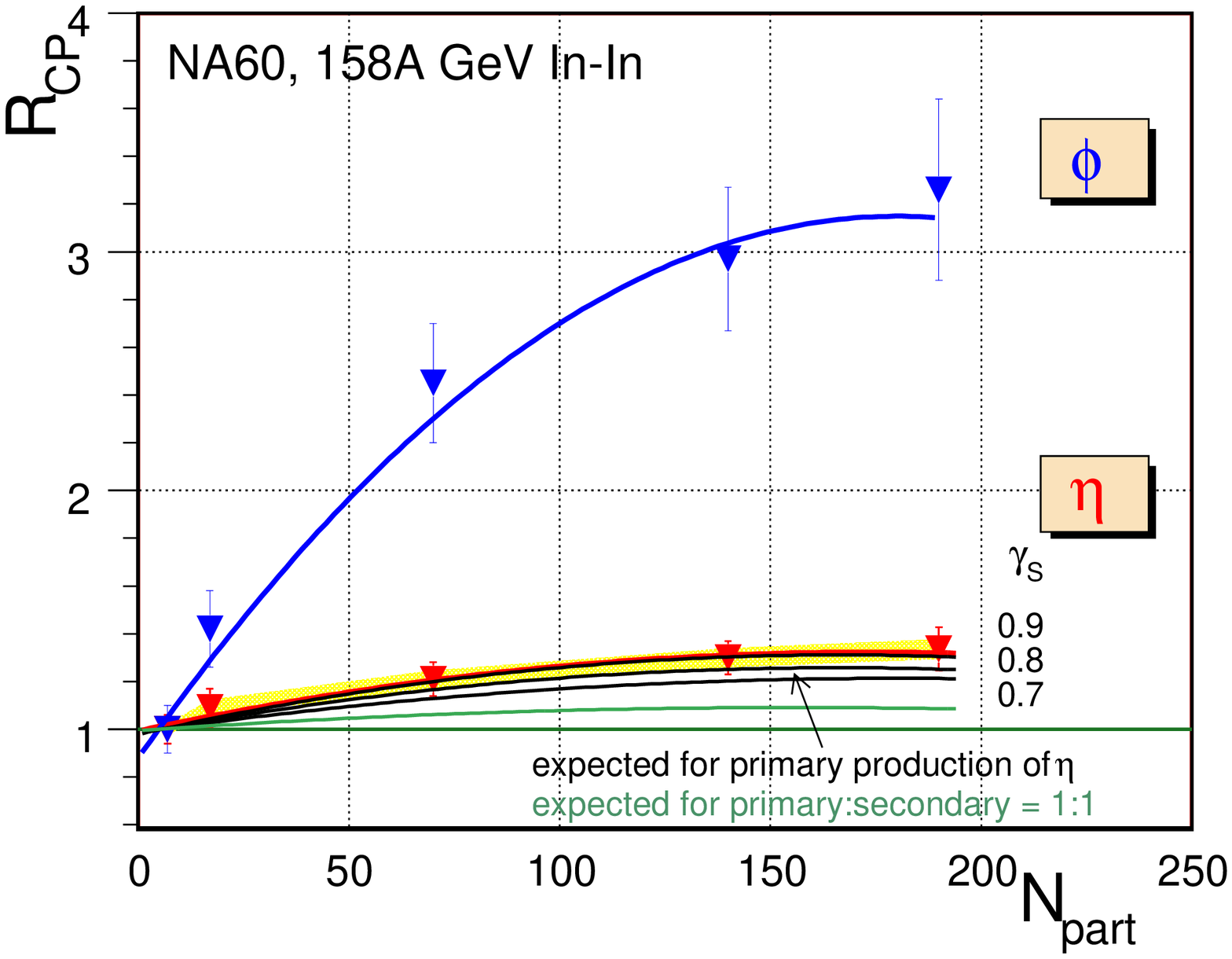}
  \caption{$R_{CP}$ for the $\eta$ compared to the one for the $\phi$ and 
the expectation for primary production only.}
  \label{fig:RCP}
\end{minipage}
\end{figure}

In fig.\ref{fig:phimult} the $\phi$ enhancement factor, expressed
as the ratio $\left < \phi \right > /N_{part}$ is plotted as a function 
of $N_{part}$, compared to the NA49 results in several collision
systems~\cite{NA49mult},~\cite{na49phi2}. The value in central In-In collisions 
exceeds the corresponding NA49 one in Pb-Pb. A direct 
comparison with NA50 requires an extrapolation of the NA50 
measurements to low $p_T$, and is difficult due to the discrepancies
in the observed inverse slopes. Two extreme cases 
with $T_{eff}=220$~MeV and $T_{eff}=300$~MeV were considered. The 
corresponding results largely exceed the NA60 values even if 
the higher $T_{eff}$ value is considered. 

To conclude, the $\phi$ puzzle is not yet solved. The forthcoming 
NA60 measurement in the $\phi \rightarrow KK$ channel may be 
decisive to clarify the picture. 

\section{$\eta$ enhancement}
The $\eta$ meson, carrying a strangeness content of $\sim$40$\%$, is
expected to be strangeness-enhanced in heavy ion collisions. 
NA60 detects the $\eta$ through its Dalitz decay 
$\eta\rightarrow \mu\mu\gamma$. In fig.~\ref{fig:RCP} the $R_{CP}$ of
the $\eta$, defined as
$R_{CP}^\eta(N_{part}) = \left <\eta \right > /N_{part}/
[\left <\eta \right > /N_{part}]_{peripheral}$, is plotted as a function of 
$N_{part}$, compared to the $\phi$. An enhancement of the $\eta$  
by a factor 1.34 from most peripheral to most central collisions is observed. 
The values for the $\eta$ and 
$\phi$ mesons can be linked through the strangeness under-saturation parameter 
$\gamma_s$. According to~\cite{Becattini}, the value of $\gamma_s$ 
is $\sim 0.7$ in Si-Si and $\sim 0.9$ in Pb-Pb central collisions. 
A value of 0.8 in central In-In collisions is assumed. 
Since $\phi$ is produced only primarily, the dependence on 
centrality of $\gamma_s$ can be then extracted from the $\phi~R_{CP}$, 
according to the scaling: 
$\left < \phi \right > / N_{part1}/\left < \phi \right > / N_{part2} = 
\gamma_s^2(N_{part1} )/ \gamma_s^2(N_{part2})$. 
The $\eta~R_{CP}$ can then be calculated under the hypothesis
that $\eta$ is only primary produced: in this case 
$\eta / N_{part} \propto 0.4 \gamma_s^2(N_{part}) + 0.6$. 
The result of the calculation is plotted in fig.~\ref{fig:RCP}, assuming 
values for $\gamma_s$ that range from 0.7 to 0.9 in central collisions. 
It should be noted that the results agree with the expectations if 
only primary production is assumed, in contrast to Statistical Models 
where a large fraction of the $\eta$ is produced from secondaries
without strangeness~\cite{BecPriv}. For comparison, if the ratio
between primary and secondary production (the latter coming
mainly from the decays of $\pi_1(1400)$ and $a_0$ which are 
not $\gamma_s$ suppressed) would be 1:1, the expected 
enhancement from peripheral to central collisions would be $\sim 1.1$. 

\section{Results in the intermediate mass region}
The dimuon intermediate mass region (IMR), located between the $\phi$ 
and the $J/\psi$ peaks, is expected to be well suited for the search of thermal
dimuons, due to its relative production yield with respect to the other
expected contributions in this region.
In p-A collisions, the IMR region is described as a superposition of the Drell-Yan
and open charm sources~\cite{imr-pa}. 
The NA38, NA50 and HELIOS-3 experiments
observed an excess in the IMR with respect to the extrapolation of these 
sources in S-U and Pb-Pb collisions~\cite{imr-pa},~\cite{imr-hi}. 
The mass spectrum of the excess
has the same shape as the open charm. Possible explanations of this
excess are the production of thermal
dimuons or the enhancement of the charm cross section. 
In the former case, the excess would have a prompt origin, 
while in the latter the muons coming from charmed meson decays
would be off-vertex. Due to limited vertex capabilities, it was not possible
to distinguish between the two scenarios. 

Thanks to its ability to measure the muon offset, NA60 can distinguish, on a 
statistical basis, between prompt and off-vertex dimuons. The resolution of 
the offset distance between the muon tracks and the vertex, estimated 
using muon pairs from $J/\psi$ decay, is 37~$\mu$m in the x coordinate
(bending plane) and 45~$\mu$m in y. Since the offset resolution of the 
matched tracks is affected by multiple scattering and energy loss, a 
weighted offset is used in the analysis: 
$\Delta_\mu=\sqrt{\Delta x^2V_{xx}^{-1} + \Delta y^2V_{yy}^{-1}
+2\Delta y\Delta yV_{xy}^{-1}}$
where $V^{-1}$ is the inverse error matrix, accounting for the uncertainties
of the vertex fit and of the muon kinematics, $\Delta x$ and $\Delta y$ are the
differences between the vertex coordinates and those of the extrapolated
muon track at $z=z_{vert}$. The dimuon offset is defined
as $\Delta_{\mu\mu}=\sqrt{(\Delta_{\mu1}^2 + \Delta_{\mu2}^2)/2}$.

NA60 observes an excess in the mass region $1.16<M<2.56$~GeV/$c^2$
with respect to the expected sources, that increases from peripheral to
central collision faster than $N_{part}$ and Drell-Yan.  

\begin{figure}[htb]
\begin{minipage}[t]{0.43\textwidth}
  \centering
  \includegraphics[width=0.90\textwidth]{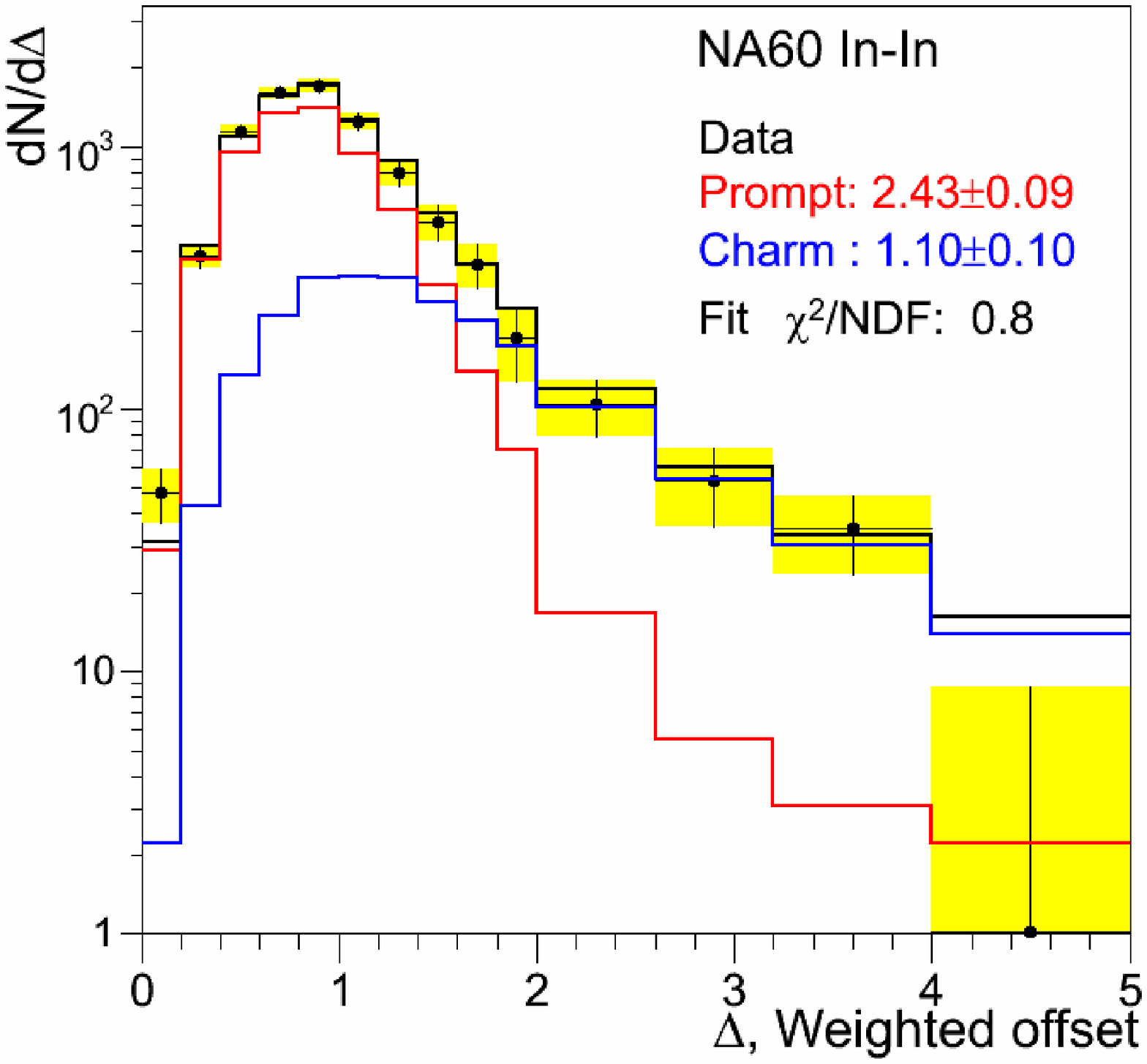}
  \caption{Dimuon weighted offset compared to the superposition of Drell-Yan
and open charm contributions.}
  \label{fig:offset}
\end{minipage}
\begin{minipage}[t]{0.57\textwidth}
  \centering
  \includegraphics[width=0.90\textwidth]{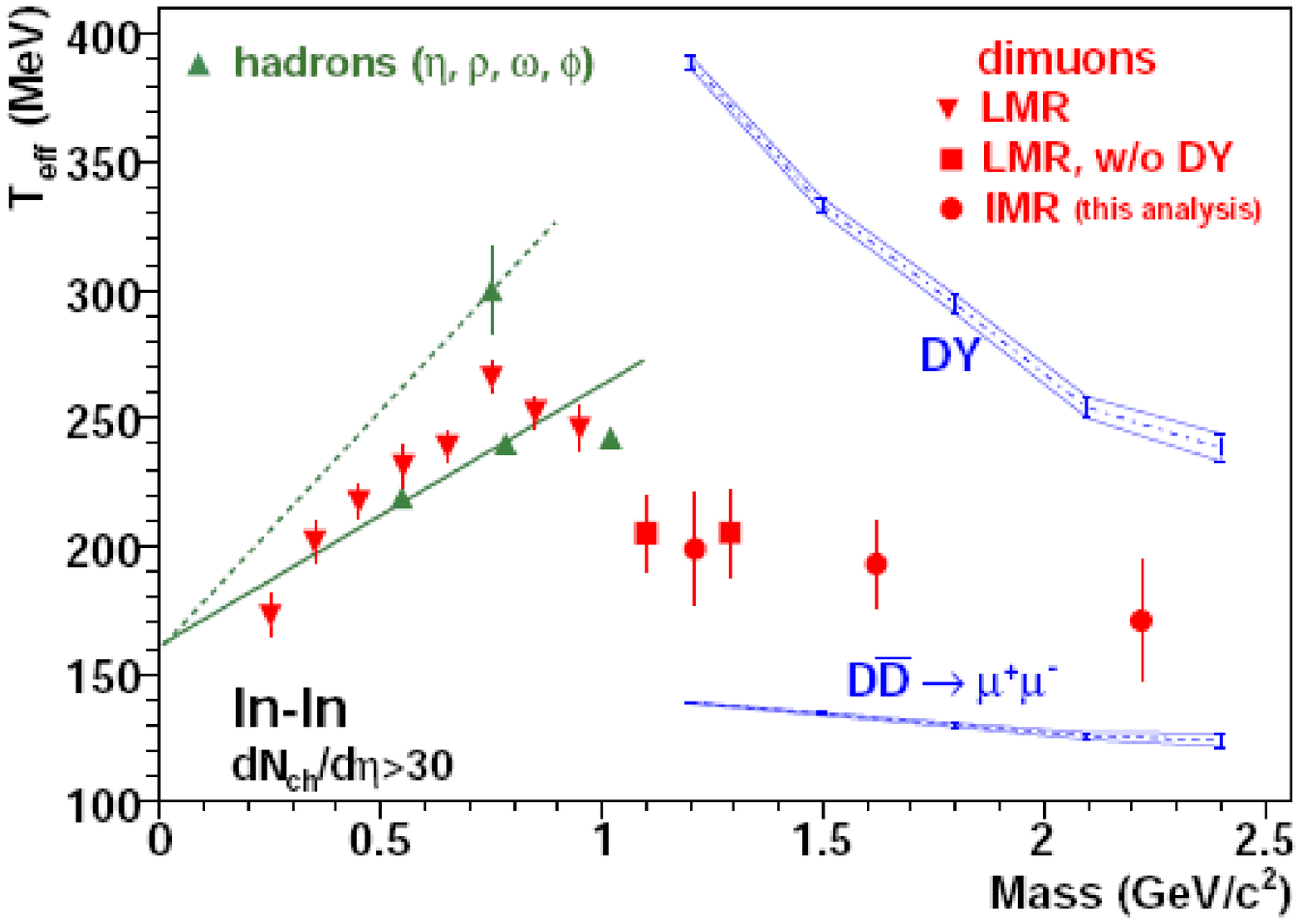}
  \caption{Inverse slope parameter versus dimuon mass. For details, 
	see~\cite{PRLflow}.}
  \label{fig:tvsmass}
\end{minipage}
\end{figure}

The dimuon offset distribution, shown in fig.\ref{fig:offset} for 
$1.16<M<2.56$~GeV/$c^2$, is compared to the corresponding 
distributions for prompt dimuons and open charm, scaled with respect
to extrapolations from p-A data. It can be noticed that, in order to 
describe the data, 
the prompt dimuon source must be scaled by a factor 2.43 with respect
to the expected Drell-Yan, while no enhancement is observed for the
open charm, within errors. Therefore, the excess can be ascribed
to a prompt source.

In order to study the properties of the excess, the expected Drell-Yan and
open charm contributions were subtracted to the data. The resulting excess
mass distribution is not reproduced by the Drell-Yan process, being much
flatter. The $p_T$ spectrum of the excess is significantly 
softer than the Drell-Yan one. Moreover, the excess $p_T$ spectra extracted 
from different mass windows in the IMR are clearly different, while for the
Drell-Yan pairs mass and $p_T$ spectra factorize, and are thus 
independent. More details on the analysis and results are reported in~\cite{imrNA60}.
The properties of the observed prompt excess, being different from the 
Drell-Yan ones in many observables, suggest an interpretation as 
thermal radiation. 

The inverse $m_T$ slope extracted for three mass windows in the IMR, is
plotted in fig.~\ref{fig:tvsmass} as a function of the mass, together 
with the corresponding values
obtained in the low mass region~\cite{PRLflow}.         
The inverse slopes for the Drell-Yan and open
charm contributions are shown for comparison.
Below 1~GeV/$c^2$, the inverse slope parameters monotonically rise with mass 
from the dimuon threshold where $T\sim 180$~MeV, up to the nominal pole 
of the $\rho$ meson, where $T\sim 250$~MeV. This is followed by a sudden 
decline to $\sim 190$~MeV in the IMR. The initial rise is consistent with the 
expectation for radial flow of a hadronic source (here 
$\pi^+\pi^-\rightarrow \rho \rightarrow \mu\mu$). The jump of $\sim50$~MeV
observed beyond 1~GeV/$c^2$, down to a low-flow situation, is 
extremely hard to reconcile with emission sources which continue to be 
of dominantly hadronic origin. Rather, the sudden loss of flow is most 
naturally explained as a transition to a qualitatively different source, 
implying dominantly early, i.e. partonic processes like $q \bar q\rightarrow \mu\mu$,
for which flow has not yet built up. 

\section{Summary}
The NA60 experiment measured dimuon production in In-In collisions at the 
CERN SPS. The $\phi$ yield observed in the leptonic channel is higher than 
the one observed by NA49 in the hadronic channel in Pb-Pb collisions. 
The dependence of the inverse slope on $p_T$ suggests that the difference 
between NA49 and NA50 is larger than expected from radial flow. 
The forthcoming NA60 measurement in the $\phi \rightarrow KK$ 
channel may be decisive to clarify the picture. 

The $\eta$ meson shows an enhancement of a factor 1.34 from peripheral
to central collisions. This enhancement, compared to the $\phi$ one, 
is consistent with the expectation in the hypothesis that the $\eta$ is only 
primarily produced. 

In the intermediate mass region, an excess of prompt origin is observed. 
The Drell-Yan process can not reproduce the properties of the excess. 
The value of $T_{eff}$, significantly smaller than in the low mass region, 
suggests an emission from the partonic stage, when the transverse flow
has not yet built up.

\end{document}